\documentstyle[aps,pre,preprint]{revtex}
\begin{document}

\def\wig#1{\mathrel{\hbox{\hbox to 0pt{%
\lower.5ex\hbox{$\sim$}\hss}\raise.4ex\hbox{$#1$}}}}
\def\lsim{\wig <}
\def\gsim{\wig >}

\draft
\tighten
\twocolumn[\hsize\textwidth\columnwidth\hsize\csname @twocolumnfalse\endcsname

\bigskip

\title {Structural Relaxation, Self Diffusion and Kinetic Heterogeneity
in the Two Dimensional Lattice Coulomb Gas }

\author{Sung Jong Lee$^1$,  Bongsoo Kim$^2$, and Jong-Rim Lee$^3$}

\address{ $^{1}$ Department of Physics,   
The University of Suwon, Hwasung-Gun, Kyunggi-Do 445-743, Korea }

\address{ $^{2}$ Department of Physics,   
  Changwon National University, Changwon 641-773, Korea }

\address{ $^{3}$ Division of Electronics, Computer and Telecommunications
Engineering, Pukyong National University, Pusan, 608-737, Korea }
\maketitle
\begin{abstract}

We present Monte Carlo simulation results on the equilibrium 
relaxation dynamics in the two dimensional lattice Coulomb gas, 
where finite fraction $f$ of 
the lattice sites are occupied by positive charges.
In the case of high order rational values of $f$ close to the 
irrational number $1-g$ ($g\equiv(\sqrt{5} -1)/2$ is the golden mean),
we find that the system exhibits, for wide range of temperatures above
the first-order transition, a glassy  behavior
resembling the primary relaxation of supercooled liquids. 
Single particle diffusion and structural relaxation show that there
exists a breakdown of proportionality between the time scale
of diffusion and that of structural relaxation analogous to
the violation of the Stokes-Einstein relation in supercooled liquids. 
Suitably defined dynamic cooperativity is calculated to 
exhibit the characteristic nature of dynamic heterogeneity present 
in the system. 

\end{abstract}

\pacs{PACS No.:\ 64.70.Pf, 64.60.Cn}

\vskip2pc]
\narrowtext


\section{Introduction}

The dynamics of supercooled liquids approaching the glass transition
remains one of the most fundamental problems in condensed matter physics 
\cite{review}. Some of the prominent dynamic features 
in supercooled liquids include  the enormous increase in relaxation
time scale with lowering temperature, 
 and the nonexponential relaxation in the response to an
external perturbation.
In addition to these features, 
an anomaly in transport properties such as breakdown of
the Stokes-Einstein (SE) relation in highly supercooled liquids
has been observed in experiments \cite{fujara} and simulations
\cite{mount,harrowell,onuki}. 
Although there exist some theoretical attempts \cite{stilinger,tarkiv,liuopp,bb,xia},
the underlying microscopic mechanism for the violation of the SE relation 
is not well understood.
Recently, there have been many experimental and  simulational studies 
of supercooled liquids that demonstrate the existence of kinetic 
heterogeneity  which  was often invoked to explain 
the origin of the non-exponential relaxation as well as the 
breakdown of the SE relation \cite{sillescu}.

In relation to these questions on microscopic slow dynamic features in
supercooled liquids, we deemed it worthwhile to investigate whether 
similar dynamic features can be found in simpler lattice spin systems or
lattice gas systems. 
In this work, we show that the aforementioned features of supercooled 
liquids, {\it i.e.}, slowing-down, non-exponential relaxation and
the (analogue) of the breakdown of the SE relation, 
are also observed in a two-dimensional (2D) lattice Coulomb gas (LCG) system.
We also find that the relaxation of the system
exhibits a spontaneous appearance of spatial heterogeneity, which we 
argue is the underlying cause for the non-exponential relaxation and the 
breakdown of the SE relation.  

In recent years, there have been some efforts to find glassy
dynamic features in the lattice spin systems with nonrandom interactions
\cite{bulbul}. 
One of the well-known examples of disorder-free lattice model
system is uniformly frustrated XY (UFXY) models in two dimensions, which 
serve as a model for two dimensional arrays of Josephson junctions under 
the influence of uniform transverse magnetic fields.      
Recent work \cite{kl} has shown that, irrespective of the true 
nature of the low temperature phase of this system,
the equilibrium dynamics of UFXY
model in the intermediate range of the temperature for frustration
parameter $f$ near $1-g \equiv (3-\sqrt{5})/2 \simeq 0.382$,
exhibits a close analogy to that of supercooled liquids.
Both spin and vortex dynamics show stretched exponential relaxations
with temperature dependent stretched exponents.
In order to investigate the dynamics of this system in more detail,
we attempted to calculate the self diffusion
properties of vortices. However, it turned out to be numerically 
ambiguous and tricky to trace the trajectories of individual vortices. 
This is beacause individual vortex around a plaquette is defined in terms
of phase angles and one probes the movement of individual vortices 
not directly but only indirectly through changes of phases, which, at
times especially when multi-vortex motion occurs, makes it ambiguous to 
determine the original position of a vortex corresponding to a new 
neighboring vortex.        

One way to overcome this difficulty was to  map the 
UFXY model onto a LCG via Villain transformation \cite{villain}, 
where the positive charges in the LCG correspond to the positive current
vortices in UFXY models. One can readily probe the diffusive dynamics
of charges without ambiguity in the LCG unlike the case of  UFXY model. 
Hence we can investigate both the structural relaxation
dynamics and self diffusion dynamics of individual vortices in LCG with 
relative ease. 

With this advantage,  we have numerically investigated the equilibrium 
relaxation dynamics 
and diffusion characteristics of LCG with charge density factor $f$ 
near $1-g \simeq 0.382$. 
We observe that for some range of temperatures above the first order 
transition, the equilibrium relaxation exhibits slow dynamic features such as 
stretched exponential relaxation and breakdown of proportionality between 
diffusive time scale and structural relaxation time scale.

It was a common belief that the 2D UFXY model and the corresponding LCG
belong to the same universality class with essentially the same 
phase transition properties, ground state symmetry, for example. However,
recent work on LCG by Gupta, Teitel, and Gingras (GTG)\cite{gtg} and also
another work on UFXY model by Denniston and Tang (DT)\cite{dt}, showed 
that there exist some difference between the two model systems especially 
in the case of dense frustration.  Both model systems exhibit first order
transition but the low temperature vortex configurations in UFXY models 
are different from the charge configurations of the corresponding LCG
for $f$ near $1-g \simeq 0.382$.
The underlying cause for this breakdown of Villain approximation in 
the limit of dense frustration is not known, but probably it is 
related to application of spin-wave integration to systems
having many metastable states with similar energies, that may cause neglect
of multi-vortex correlations.

Special interest has been given to the case of $f$ approaching $1-g$ 
\cite{halsey,mychoi}. 
Consider a system where $f$ equals $p_0 /q_0$ ($p_0$ and $q_0$
are relative primes) which is a rational approximant to $1-g$.
Here, in the case of UFXY model, DT argues that 
the low temperature vortex configuration has lattice periodicity which is 
of order $q_{0}^{2}$,  {\it i.e.}, much larger than $q_0$.
On the other hand, in the case of LCG, GTG \cite{gtg} showed, 
via Monte Carlo (MC) simulations, 
that the low temperature charge configurations are characterized by 
arrangements of diagonal stripes that are either completely filled,
completely empty, or partially filled with charges  that are quite different
from those vortex configurations in the corresponding UFXY model. 
However, GTG did not enumerate the exact patterns of low temperature
charge configurations (such as spatial periodicity) for general cases of
dense charge filling.
In this work, we find that, for the values of $f$ between $1/3$ and $2/5$,
there exist a simple regularity in the low temperature charge configuration
which consists of periodic arrangements of combinations of two out 
of three types of striped charge patterns (see Section III).  

For wide range of quenching temperatures above the first order transition
$T_c$, the equilibrium relaxation 
continues to slow down with lowering temperature, and the form of the relxations
are characterized by the stretched exponential with temperature-dependent 
exponents. Moreover, we observe that the model exhibits a separation of 
the two  characteristic time scales, i.e., the time scale of single 
particle diffusion and that of structural relaxation.
This feature is quite analogous to the breakdown of the 
SE relation observed in supercooled liquids.
Stretched exponential relaxation is
observed to be accompanied by interesting dynamic heterogeneity in the
system.  It appears that the kinetic heterogeneity  
is the underlying reason for both the stretched exponetial relaxation
and the separation of the relaxation and diffusion time scales.

 A convenient measure for dynamic heterogeneity is the so called 
dynamic cooperativity \cite{doliwa} of the particle motions.
This measures reduction of the effective degrees of
freedom. One interesting result from our simulations is that the 
magnitude of the velocity (or displacement vector) exhibits strong 
increase in cooperativity of the particle motions.
On the other hand, the displacement vector itself shows  
cooperativity a little smaller than unity due to anti-correlations
in the direction of particle motions. 
This means that the system can be divided into highly mobile regions 
and relatively inert regions resulting in highly inhomogeneous 
local mobility distribution. However, there is no macroscopic flow of
particles that will generate long range positive correlations between 
the directions of flows of particles.

When quenched to a temperature below $T_c$, 
the system is always found to undergo phase ordering via slow coarsening 
processes.  The system therefore does not remain in a supercooled state.
Rather it becomes slowly crystalized.
It should be emphasized that in this system it is the relaxation  
for the temperatures {\em above} $T_c$ that exhibits slow dynamic behavior
which shares some common features with that of supercooled liquids.

\section{Model and Simulation Methods}
 
General 2D LCG \cite{teitel} is described by the following Hamiltonian 
that can be mapped from UFXY model by means of 
Villain transformation \cite{villain},

\begin{equation}
{\cal H}_{CG}={1 \over 2}\sum_{ij}Q_iG(r_{ij})Q_j \label{eq:Hcg}
\end{equation}
where $r_{ij}$ is the distance between the sites $i$ and $j$,
and  the magnitude of charge $Q_i$ at site $i$ can take either $1-f$ or $-f$ 
, where $f$ corresponds to the frustration parameter
in the related XY models. 
Charge neutrality condition $\sum_iQ_i=0$ implies that the number density of 
the positive charges is equal to $f$. 
As was mentioned above, we can thus view the system as a lattice gas 
of $N \cdot f$ charges of unit magnitude upon uniform negative 
background charges of charge density $-f$ ($N= L^2$ is the total size
of the system with the linear dimension $L$).  
The lattice Green's function $G(r_{ij})$  solves the equation

\begin{equation}
(\Delta^2 - { 1 \over {\lambda^2}}) G(r_{ij})=-2\pi\delta_{r_{ij},0}
\end{equation}
where $\Delta^2$ is the discrete lattice Laplacian and $\lambda$ is 
the screening length which, in normal case of no screening, is set to 
an infinity. For the case of usual Villain transformation of UFXY model,
we have $\lambda = \infty$. But it is included in this equation for generality. 
Since, in this work, we restrict our attention to only square 
lattice with periodic boundary conditions, $G(r)$ is given by 

\begin{equation}
G(\vec{r})={\pi \over N} \sum_{\vec k \neq 0} { e^{i\bf{k} \cdot r}-1 \over 
{2 - \cos k_x -\cos k_y + 1/\lambda^2} } ,
\end{equation}
where $\bf{k}$ are the allowed wave vectors with $k_{\mu} = (2 \pi n_{\mu}/ L)$,
with $n_{\mu} = 0, 1, \dots, L-1 $.
In the case of infinite screening length, for large separation $r$,
 one gets $G(\vec r)\simeq -\ln r$ \cite{jrl}. In this work, we consider
the limiting case of $ \lambda \rightarrow \infty $ only.

In our MC simulations, the initial disordered random configuration 
is updated according to the standard Metropolis algorithm
by selecting a positive charge at random and moving it over to 
one of the {\em nearest neighbor (NN)} or {\em next nearest neighbor (NNN)} 
sites \cite{gtg}.  We find that this {\em NNN} hopping algorithm is 
particularly effective in terms of simulation time as compared with {\em NN}
hopping alone, as was emphasized in \cite{gtg}. Moreover, at low temperature, 
{\em NN} hopping alone presented severe energy barriers to the motions of 
charges in the case of relatively dense Coulomb gas, {\it i.e.}, $f$ 
approximately larger than $1/3$.

The presented results are averages over $100 \sim 500$
different random initial configurations depending on the temperature.
In order to ensure that equilibration is achieved, we calculate 
the two-time charge density autocorrelation function and locate the
waiting time beyond which the autocorrelation function no longer depends
on the waiting time.    
As for the values of the charge density parameter $f$, we chose 
$f = 55/144 \simeq 0.3819$ that is close to $f=1-g$, and  
square lattices of linear size $L=36$ is chosen with periodic boundary 
conditions. This value of $f$ is chosen as a 
simple rational value that satisfies the two conditions of both  being 
close to $1-g$ and being commensurate with the lattice periodicity $12$ 
as explained in Section III.   
We found that qualitative features of relaxation dynamics are 
the same for other nearby values of the frustration $f$. 

\section{Simulation Results and Discussions}

\subsection{First order transition and low temperature configuration}
  
We first discuss the equilibrium phase transition and charge configuration
of the system.
As was first shown by GTG, we also find that there exist a first order
transition in LCG with $f$ near $1-g$. Figure~1 shows temporal snapshots 
of charge configurations evolving from disordered state into ordered 
configuration after being quenched to a temperature $T=0.026$. 
First order nature of the phase transition can easily be confirmed
by enumerating the histogram 
of energies $P(E)$ near the transition temperature \cite{sw}. $P(E)$ is 
obtained by counting the occurrences of energies for each of the equally
spaced energy bins while performing the equilibrium Monte Carlo simulations 
(via simple Metropolis algorithm). 
For a system with first order transition, the energy histogram $P(E)$ 
becomes bimodal near the transition temperature corresponding to a mixture
of ordered state (with lower energy) and a disordered state 
(with higher energy). The transition temperature $T_c$ can be determined by 
locating the temperature where the subareas under the two peaks
are equal.  Figure~2 shows two histograms near the transition temperature,
where we could estimate the transition temperature approximately as 
$T_c \simeq 0.0316$. Since we did not attempt a detaled analysis (including
a finite size scaling) of the histogram, we think that  this estimate value 
of the transition temperature should not be considered too seriously for 
its precision. 

We find empirically that there exist a simple regularity in the low 
temperature charge configuration in LCG (Fig.~3). For the case of values of 
$f$ in the range $1/3 \leq f \leq 2/5$, it is found that the low temperature
configuration becomes quasi-one-dimensional with periodic striped patterns.
In the cases of $f=1/3$ and $f=2/5$ the ground state configurations are 
identical to the low temperature vortex configurations in 
the UFXY model. However, for values of $f$ in between $1/3$ and $2/5$, 
the low temperature patterns are found to be,
unlike the case of corresponding UFXY model, consisting of periodic
arrangements of combinations of two out of three types of striped charge
patterns as follows.

First component pattern (type I pattern) is a sequence of three diagonals 
which are {\em empty, filled}, and {\em empty} respectively (that may be 
denoted by {\bf (010)} in our notation where {\bf 1} refers to a filled 
diagonal and {\bf 0} refers to an empty diagonal). In other words, it is 
a pattern with single 
isolated diagonal filled with charges, that is neighbored by empty 
diagonals on both sides. 
Repetition of this pattern alone produces the ground state configuration for 
the case of $f=1/3$ with spatial periodicity three.  

Second component pattern (type II pattern) consists of a sequence of five 
diagonals that are  {\em empty, filled, empty, filled}, and {\em  empty } 
respectively, or 
{\bf (01010)} in our notation. This may be termed as a double filled diagonal 
because two filled diagonals are positioned in parallel at second neighbor. 
This forms the basis
of the ground state configuration for the case of $f=2/5$ with 
lattice periodicity five.   

Lastly, the third component pattern (type III pattern) consists of a sequence
of seven diagonals that are sequentially {\em empty, filled, empty, partially 
filled, empty, filled}, and {\em  empty } {\it i.e.},  {\bf (010p010)} in our 
notation where {\bf p} refers to a partially filled diagonal where only part 
of the diagonal sites are occupied by positive charges. This pattern is 
essentially a partially filled diagonal enveloped by two filled diagonals 
on both sides at second neighbor diagonal position, which may be termed as
a {\em channel} structure. This can form a basis with spatial lattice
periodicity seven.

We leave the detailed description of the low temperature charge patterns 
for the full range of $f$ values between $1/3$ and $2/5$ to the
forthcoming publication \cite{config_cg}. And we describe in this work 
the low temperature ordered patterns for values of $f$ around $1-g$ only.
Near the value of the filling ratio $f=1-g \simeq 0.382$, 
we find that, among the three patterns above, only two types (type II and
type III patterns) participate in the stable charge configurations with the
resulting spatial lattice periodicity depending on the combination of the 
two component patterns.

We find that there exist a value $f=f_c \simeq 0.381$ which separates 
two regimes with distinct low temperature striped patterns. 
For values of $f$ in 
the range $0.36 \lsim f \lsim 0.381$, the stable striped patterns 
turn out to have periodicity $l_p = 7$ which consists of simple
repetitions of channel structures (type III pattern). 
Note that this periodicity seven refers to the periodicity of the filled
diagonals only (neglecting the true periodicity including the charge 
configurations within the partially filled diagonals).

On the other hand, for values of $f$ in the range $0.381 \lsim f \lsim 0.39$, 
the stable configuration exhibits a periodicity $l_p=12$, which consists of 
double filled diagonals (type II) and channels (type III) alternatingly 
placed.
As the value of $f$ continuously increases within the two regimes (in the
above), the system in the low temperature stable configuration simply 
adjusts itself by accomodating the extra charges into the partially filled
diagonal channels and thereby changing the charge filling within the channels.
The dividing value of $f=f_c \simeq 0.381$  between the two regimes appears
to correspond to the value
$8/21$ in which case the partially filled diagonals have filling density
exactly equal to $2/3$.
Our simulations show that the filling density $2/3$ inside the partially filled
diagonal plays as a stability limit for the channel structures. 
Beyond this limit, electrostatic instability probably 
begins to set in, and rearrangement of the whole charge configuration occurs
in order to form a new stable ordered patterns. 
As was also argued by GTG, in general, at much lower temperature 
$T_p$ (below $T_c$) the charges within the partially filled channels are 
expected to exhibit ordering, which would depend sensitively on rationality of
the exact filling ratio of charges inside the partially filled diagonals.

An important aspect of our simulations is that one has to choose the 
lattice size appropriately in order to match the periodicity of the true low 
temperature configuration in the thermodynamic limit. If, otherwise, one 
chooses a lattice size that is incommensurate with the periodicity 
(of striped patterns), then one ends up with defective charge configurations 
with patches of local ground state configurations. 
We think that this is probably why GTG got two 
different equilibrium configurations when two different lattice sizes $L=26$ 
and $L=52$ are used for $f=5/13$ since these $L$'s turn out to be
incommensurate with the true periodicity $l_p =12$. 

When the screening length $\lambda$ is finite,
then we find the low temperature configuration becomes 
different from the case of no screening ($\lambda \rightarrow \infty$)
in such a way that the partially filled diagonals gets rarer.
The influence of the screening effect on the statics and the relaxation dynamics
needs further study.

\subsection{Equilibrium relaxation dynamics}

We now discuss the equilibrium relaxation dynamics of the model
above first order transition.
In order to probe the structural relaxation of charges, 
we measured the on-site charge autocorrelation functions, 

\begin{equation}
C(t)=\langle \sum_{i=1}^{N} Q_i(0)Q_i(t) \rangle / N,
\end{equation}
where  the bracket $< \cdots >$ represents an average over different random 
initial configurations. 

Shown in Fig.~4a is the on-site charge autocorrelation 
function $C(t)$ for temperatures from $T=0.1$ down to $T=0.033$.
From this figure, we observe a slowing down in the structural relaxation for
this temperature range. 
One can extract a characteristic time scale $\tau(T)$ which, for example,  is
defined as $C(t=\tau(T))=1/e$ for each temperature $T$.  As Fig.~4b clearly 
shows, the temperature dependence of the relaxation time exhibits a 
non-Arrhenius behavior.  We also checked whether the so-called 
time-temperature superposition holds for the above autocorrelation 
functions, which is shown in Fig.~4c. We clearly see that time-temperature 
superposition is systematically broken by the autocorrelation functions.
This is consistent with the fact that the stretched exponents have 
dependence on temperature as is shown just below. 

 
We find that the relaxation pattern of the correlation function $C(t)$
can be  characterized by a power law relaxation $C(t)=1-A t^{b(T)}$
(known as the von Schweidler relaxation) 
in the early time regime and a stretched exponential relaxation
$C(t)=C_0(T) \exp(-A't^{\beta(T)})$ in the late time regime. However, 
as the temperature gets higher, the regime of validity for early time
power law relaxation was significantly reduced and we could better fit
the early time relaxation with another stretched exponential form
$C(t)=\exp(-A''t^{b'(T)})$. Of course for low temperature regime, we
could get $b(T) \simeq b'(T)$.

Fig.~4d shows the temperature dependence of the fitted exponents.
We see that non-exponentiality increases as the temperature decreases.
These results clearly indicate that the equilibrium relaxation
in the 2D LCG above $T_c$
closely resembles the primary relaxation of typical fragile liquids.

One of the main characteristic features of the single particle dynamics is 
described by the mean square displacement $\langle (\Delta \vec r)^2 \rangle$,
which is defined as

\begin{equation}
\langle (\Delta \vec r)^2 \rangle = \langle \sum_{j=1}^{N_Q}
 (\vec{r}_j(t)-\vec{r}_j(0))^2 \rangle / N_Q,
\end{equation}
where $\vec{r}_j(t)$ is the position vector of the $j$-th charge at time $t$
and $N_Q$ the total number of charges.
Figure 5 shows $ \langle (\Delta \vec r)^2 \rangle $ for various 
temperatures.  It exhibits an early time subdiffusive regime and 
crosses over into late time diffusive regime. Early time subdiffusive 
behavior is thought to be 
coming from local frustrated motions of charges before reaching an average
displacement of unit lattice spacing.
To test the proportionality of the two time scales, the structural relaxation
time scale $\tau$ and the diffusion time scale $D^{-1}$, we plot the 
temperature dependence of the product $4 D\tau$ in Fig.~6.
Here, we clearly see that the breakdown of the proportionality between
the two time scales is observed for wide range of temperatures below 
$T=0.1$ and becomes stronger as the temperature is lowered.
This separation of the two time scales is 
due to the weaker temperature dependence of the diffusion coefficient.
That is, diffusion is relatively enhanced at lower temperatures.
This is quite analogous to the violation of the SE relation ($D=T/a \eta$, 
where $a$ is a molecular length and $\eta$ is the viscosity of the liquid) 
observed in experiments on supercooled liquids  \cite{fujara}. 
Here, we mention that there exists a correlation between the increase
of non-exponentiality (as the temperature is lowered) and the increase 
of the product $4 D\tau$ at low temperatures \cite{ediger}.

If we suppose that there exists a single dominant relaxation mode in
the system (and hence one relaxation time scale $\tau$), then we
would obtain a simple exponential behavior for the relaxation function
$C(t) \sim e^{-t/\tau}$. On the other hand, if the system consists
of many regions with different relaxation times, then the
relaxation function would be roughly some superposition of
exponentials with a broad distribution of relaxation times,
which would be in general not expressible in a simple exponential 
form, but in stretched exponetial form or other more complicated
forms.  

The fact that there exists a breakdown of proportionality between 
$\tau$ and $D^{-1}$ can be interpreted in the following way that 
invokes dynamic heterogeneity. As the temperature is lowered, the 
system consists of many regions with different relaxation time
that comes from different local mobilities. We can easily see that
the structural relaxation time is dominated by the least 
mobile regions, that is, by the regions with the longest relaxation 
time. On the contrary, the average (long time) diffusion 
characteristics is dominated  by the most mobile regions.  
In other words, the structural relaxation function and the
self-diffusion function, respectively, are probing more or less
opposite aspects of the relaxation behavior of 
the system. For an extreme example, one can imagine a system where
half of the whole system is frozen (no motion of the component
particles) while the remaining half of the system has finite 
relaxation time with uniformly distributed mobile particles. For
this system, the structural relaxation time $\tau$ would be infinite
due to the frozen half of system, but inverse of the average 
diffusion constant $D^{-1}$ is finite due to the mobile part of
the system, leading to an extreme breakdown of SE relation.
Above simulation result, thus, can be interpreted as an evidence
pointing toward the existence of a kinetic heterogeneity
in the relaxation dynamics and mobility of the system. 

In fact, the kinetic heterogeneity can be visualized in our system.
Typical charge configuration at $T=0.033$, as shown in Fig.~7, 
exhibits local striped patterns (ordered domains)
and interfacial regions due to mismatch between adjacent domains. 
For a fixed quenching temperature, the average size of these local domains  
reaches a certain length scale when the system equilibrates.  After
equilibration, 
the system structurally rearranges itself going from one configuration
to another with local domains of similar length scale.
Figure 8 shows the trajectories of moving positive charges over a 
time interval of 500 MC steps for $T=0.033$ (corresponding to Fig.~7). 
We can see that there exist local regions with actively moving charges 
and other regions with relatively immobile charges.
Among the active regions, we can find those charges moving along 
partially filled diagonal channels. We also find some extended interfacial 
regions where no discernible local order can be identified, that exhibit
relatively high mobility.
Enhancement of particle diffusion is probably due to the motions of charges
along the partially filled diagonals as well as those fluidized motions in 
the extended interfacial regions. 
These fastly moving regions in  surroundings of  very slowly moving 
regions offer a specific
example for spatial heterogeneity in  glassy systems
\cite{harrowell,onuki}, 
which was often thought of as the physical mechanism for breakdown of the 
SE relation. 

One simple way to quantify the degree of dynamic heterogeneity 
directly from the local motions of particles is to calculate 
the dynamic cooperativity \cite{doliwa} for one particle dynamic quantities 
such as {\it e.g.}, displacement
vectors $X_i \equiv |\vec{r}_i (t+ \Delta t) - \vec{r}_i (t) | $ 
between the time $t$ and $t+\Delta t$ for some fixed time interval 
$\Delta t$.
 We can also choose $X_i$ to be the vector displacement itself
$X_i \equiv \vec{r}_i (t+ \Delta t) - \vec{r}_i (t) $. 
If there are no correlations between the motions of particles, then 
the variations of the $X_i$'s will satisfy 

\begin{equation}
\sigma [ \sum_{i} X_i ] = \sum_{i} \sigma [X_i ],
\end{equation}
where $\sigma[x]$ denotes the mean square deviations of the 
random number $x$, $\sigma [x] \equiv \langle (x-<x>)^2 \rangle$. 
However, some correlations between the particle motions will increase
$\sigma [ \sum_{i} X_i ]$ or anti-correlations will decrease it. 
Following Doliwa and Heuer, we can define the dynamic cooperativity as
\begin{equation}
N_{X}^{coop} \equiv {{\sigma [ \sum_{i} X_i ] } \over 
{ \sum_{i} \sigma [X_i ]}}.
\end{equation}
In the case of no correlations between the motions of particles,
as in (6), we get $N_X^{coop} =1$. If there exist some positive correlated
motions between particles, we would get $N_X^{coop} > 1$, while 
anti-correlations between the motions of particles would result 
in $N_X^{coop} < 1$.
Doliwa and Heuer investigated the dynamic cooperativity of hard sphere 
systems in 2D and 3D, where they found finite cooperativity 
($N_X^{coop} > 1$) for both vector displacement and the scalar 
magnitude of the displacement, 
which is consistent with the snapshots of the particle motions in their
work. They argue that the dynamic cooperativity measures the total reduction of
degrees of freedom due to the correlations. 
Here we also studied the dynamic cooperativity of the lattice gas particles 
by calculating $N_{X}^{coop}$ for both the scalar displacement 
and the vector displacement itself.
 Interestingly, we found that the scalar 
displacement exhibited finite dynamic cooperativity (Fig.~9a), while the vector 
displacement itself showed weak anti-correlations between particles. 
as shown in Fig.~9b. In the case of scalar displacement,
the cooperativity increases at first as the time interval $\Delta t$ increases
and reaches its maximum near the $\alpha$-relaxation time scale $\tau$. 
Then it decrease back to values around unity (corresponding to no 
correlations) at large $\Delta t$. 

Contrasting features of cooperativity for our LCG system and that for 
the hard sphere systems may be interpreted as follows.
In the case of hard sphere systems near the glass transition, the packing 
density is very high and the inter-particle interaction is a short ranged
one. Therefore, the local motions of particles in hard sphere systems are 
naturally highly correlated in both its direction and magnitude due to the
continuity constraint of particles resulting in a large scale flow with
directional correlations. 

In contrast, in the case of the LCG, the density of
particles is relatively low ($f \simeq 0.38 $) as compared with the case
of hard sphere systems near the glass transition. In addition to that,  
charge motions in the LCG is driven by thermal effect.
From the snapshots of charge configurations, we see that there 
exist locally mobile regions as well as locally immobile regions. 
Locally immobile regions consist of charge configurations that are
close to the low temperature striped patterns. Mobile regions, however,
consist of charges that are agitated in random directions
due to 
the thermal effect. Thus we do not observe positive dynamic cooperativity
in vector displacement, but only the scalar displacement exhibits 
appreciable positive cooperativity due to the local regions with 
high mobilities. 
Hence, heterogeneity still exists in our lattice Coulomb gas in terms of
local mobility distribution, but unlike the case of hard sphere systems,
there is no appreciable average local flow. 

Also, we may look into the nature of the equilibrium dynamics of
the system in wave-vector space.
Figure~10 shows the structure factor $S(q) \equiv \langle |\rho_q|^2 \rangle $
at equilibrium where  $\rho_q \equiv \sum_j \exp [i 
\vec{q} \cdot \vec{r}_j ]/N $ where $q= {{2 \pi }  \over L} n $, 
$n=1,2, \cdots 2/L $. We see that the structure factor of our LCG
shows some similarity to those of dense liquids with first peak
corresponding roughly to the inverse of the average distance between
charges. Due to the lattice nature of the LCG, the wave vector has 
cutoff value at $q_{max}=\pi$ as in the figure. 

The diffusive properties of the system can be probed by  
calculating the incoherent scattering function (ISF) $F_S(q,t)$
which is defined as in our model of LCG 

\begin{equation}
F_S (q,t) \equiv \langle \sum_{j=1}^{N_Q} \exp i \vec q 
\cdot [\vec r_j(t)- \vec r_j (0)]
\rangle / N_Q,
\end{equation}
where $\vec r_j(t)$ denotes the position of $j$-th particle on 
the lattice. Due to the
 discrete lattice nature of our model sytem, we need
to consider the wave-vectors within the first Brillouin zone 
$q= {{2 \pi }  \over L} n $, $n=0,1,2, \cdots L-1 $.
Figure~11 shows the $q$-dependence of $F_S(q,t)$ at temperature $T=0.033$.
We find that the long-time behavior of $F_S(q,t)$ also can be fitted to
stretched exponential form. For low $q$, the late time $\beta$ exponents
were close to one (pure exponential relaxation) but as $q$ increases the 
exponents decreased down to $\beta \approx 0.73$ for
$q = 18 \times 2 \pi /36 $, and $T=0.033$ (Fig.~12).
As can be seen from the definition of $F_S(q,t)$, for gaussian 
distribution for the displacement vector $\Delta \vec{r}_i$, 
we would get  
\begin{equation}
 F_{G} (q,t) \equiv \langle \exp iq [\Delta r ] \rangle 
= \exp [ -{{q^2} \over 2} \langle (\Delta r )^2 \rangle ]. 
\end{equation}
Figure 13 shows that the gaussian approximation is quite good for low $q$.
That is, for long distance diffusion, the distribution gets closer to 
gaussian.  However, as $q$ becomes larger, the gaussian approximation gets
worse as shown in the figure. Similar features were reported in molecular
dynamics simulations on the dynamics of supercooled water \cite{sciortino}. 

In summary, we have shown that the 2D LCG with fractional filling of 
charges exhibits an equilibrium relaxation behavior, above first order melting 
transition,
characterized by two time-regimes of stretched exponetial form 
with temperature dependent exponents, which is quite similar 
to the primary relaxation of typical 
supercooled liquids. We found a strong deviation from proportionality between
the diffusive time scale and the structural relaxation time scale 
resembling the breakdown of SE relation in supercooled liquids.
This is accompanied by a characteristic dynamic cooperativity, where the scalar
displacement exhibits positive cooperativity while the vector displacement
shows anti-correlations leading to the vector cooperativity less than unity.  
We have identified the microscopic heterogeneous structure which is 
responsible for this phenomena.

\bigskip

We thank M. D. Ediger, P. Harrowell, K. Kawasaki and S. Teitel for discussions.
This work was supported by the Korea Research Foundation Grant
(KRF-1999-015-DP0098) (SJL, BK) and (KRF-1998-15-D00089) (JRL).

\newpage

\centerline {\bf FIGURE CAPTIONS}

\renewcommand{\theenumi}{Fig.~1}
\begin{enumerate}
\item
Snapshots of charge configuration at time steps (a) t=16 MCS, (b) t=4096 MCS,
(c) t=65536 MCS, and (d) t=1048576 MCS, for temperature $T = 0.026$ and
$f=55/144$, exhibiting coarsening toward an ordered striped state.
Positive charges are represented by filled squares.
\end{enumerate}

\renewcommand{\theenumi}{Fig.~2}
\begin{enumerate}
\item
Energy histogram near the first order transition temperature
(for $T = 0.03165$ and $T=0.0317$).
\end{enumerate}

\renewcommand{\theenumi}{Fig.~3}
\begin{enumerate}
\item
Regimes of charge patterns for the range of value of $f$ between
$1/3$ and $2/5$. See the text for details. 

\end{enumerate}

\renewcommand{\theenumi}{Fig.~4}
\begin{enumerate}
\item
     (a) The charge autocorrelation functions for temperatures 
    $T = 0.1$, $0.08$, $0.06$, $0.05$, $0.042$, $0.037$,
       $0.035$, $0.033$. 
(b) Arrhenius plot for the relaxation time ($\log (\tau)$ versus 
$1/T$). 
(c) Charge autocorrelation functions in (a) replotted in terms of
the rescaled time $t/ \tau (T)$ which shows that the time-temperature 
superposition is broken. 
(d) Temperature dependence of the $b$ and $\beta$ exponents for 
charge autocorrelation functions.
\end{enumerate}

\renewcommand{\theenumi}{Fig.~5}
\begin{enumerate}
\item
 Squared displacement $W(t)$ versus time $t$ for the same temperatures 
 as in Fig.~4a.
\end{enumerate}

\renewcommand{\theenumi}{Fig.~6}
\begin{enumerate}
\item
Comparison of the two time scales $D^{-1}$ and $\tau$
($4D \tau $ versus $T$), which clearly shows that the diffusive time 
scale increases slowly (as the temperature is lowered) as compared with
the structural relaxation time. 

\end{enumerate}
    
\renewcommand{\theenumi}{Fig.~7}
\begin{enumerate}
\item
  Typical charge configurations at $T=0.033$.
  Positive charges are represented by filled squares.
\end{enumerate}

\renewcommand{\theenumi}{Fig.~8}
\begin{enumerate}
\item
 Trajectories of moving positive charges at $T=0.033$ over a time
 interval of $500$ MC steps. Arrows indicate the directions of single charge
 motions.
\end{enumerate}

\renewcommand{\theenumi}{Fig.~9}
\begin{enumerate}
\item
Dynamic cooperativity for (a) scalar displacement and (b) vector
displacement respectively for varying time intervals at various temperatures. 
\end{enumerate}

\renewcommand{\theenumi}{Fig.~10}
\begin{enumerate}
\item
  The structure factor $S(q)$ at $T=0.033$ and $T=0.037$.
\end{enumerate}

\renewcommand{\theenumi}{Fig.~11}
\begin{enumerate}
\item
  The incoherent intermediate scattering functions at
  temperature $T = 0.033$ for various wave vectors $q$.
\end{enumerate}

\renewcommand{\theenumi}{Fig.~12}
\begin{enumerate}
\item
 $q$-dependence of the $b$ and $\beta$ exponents for  
the intermediate scattering functions at temperature $T = 0.033$.
\end{enumerate}

\renewcommand{\theenumi}{Fig.~13}
\begin{enumerate}
\item
  Comparison of the Gaussian approximations and the incoherent 
  intermediate scattering functions at
  temperature $T = 0.033$ for various wave vectors $q$. We can see that
  the gaussian approximation is worse at large wave vectors. 
\end{enumerate}


\begin{thebibliography}{99}
\bibitem{review}
M. D. Ediger, C. A. Angell, and S. R. Nagel, J. Phys. Chem. {\bf 100}, 13200 (1996);
most recent developments on the subject can be found in the collection of papers
in the conference proceedings such as J. Non-Cryst. Solids {\bf 235-237} (1998),
J. Phys. C: Condens. Matter {\bf 10A} (1999), and J. Phys. C: Condens. Matter 
{\bf 12} (2000).

\bibitem{fujara}
F. Fujara, B. Geil, H. Sillescu, and G. Fleischer, Z. Phys. B {\bf 88}, 195 (1992);
N. Menon, S. R. Nagel, and D. C. Venerus, Phys. Rev. Lett. {\bf 73}, 963 (1994);
M. T. Cicerone and M. D. Ediger, J. Chem. Phys. {\bf 104}, 7210 (1996) 
and references therein.

\bibitem{mount}
D. Thirumalai and R. D. Mountain, Phys. Rev. E {\bf 47}, 479 (1993).

\bibitem{harrowell}
D. N. Perera and P. Harrowell, Phys. Rev. Lett. {\bf 81}, 120 (1998).

\bibitem{onuki}
R. Yamamoto and A. Onuki, Phys. Rev. E  {\bf 58}, 3515 (1998); Phys. Rev. Lett. 
{\bf 81}, 4915 (1998).

\bibitem{stilinger}
J. A. Hodgdon, and F. H. Stillinger, Phys. Rev. E {\bf 48}, 207 (1993); F. H. Stillinger
and J. A. Hodgdon, {ibid}, {\bf 50}, 2064 (1994).

\bibitem{tarkiv}
G. Tarjus and D. Kivelson, J. Chem. Phys. {\bf 103}, 3071 (1995).

\bibitem{liuopp}
C. Z.-W. Liu and I. Oppenheim, Phys. Rev. E {\bf 53}, 799 (1996).

\bibitem{bb}
S. Bhattacharyya and B. Bagchi, J. Chem. Phys. {\bf 107}, 5852 (1997).

\bibitem{xia}
X. Xia and P. G. Wolynes, cond-mat/0101053.

\bibitem{sillescu}
For recent review on the heterogeneity, see
H. Sillescu, J. Non-Cryst. Solids {\bf 243}, 81 (1999) 
and references therein.

\bibitem{bulbul}
For some of the most recent works, see 
B. Chakraborty, L. Gu, and H. Yin, J. Phys.: Condens. Matter {\bf 12}, 6487 (2000);
A. Lipowski and D. A. Johnson, Phys. Rev. E {\bf 61}, 6375 (2000);
M. Swift, H. Bokil, R. D. M. Travasso, and A. J. Bray, 
Phys. Rev. B {\bf 62}, 11494 (2000).


\bibitem{kl}
B. Kim and S. J. Lee, Phys. Rev. Lett. {\bf 78}, 3709 (1997);
S. J. Lee and B. Kim, Phys. Rev. E {\bf 60}, 1503 (1999).

\bibitem{villain}
J. Villain, J. Phys. (Paris) {\bf 36}, 581 (1975);
J. V. Jos\'e, L. P. Kadanoff, S. Kirkpatrick, and D. R. Nelson, 
Phys. Rev. B {\bf 16}, 1217 (1977).

\bibitem{gtg}
P. Gupta, S. Teitel, and M. J. P. Gingras, Phys. Rev. Lett. {\bf 80}, 105 (1998).

\bibitem{dt}
C. Denniston and C. Tang, Phys. Rev. B {\bf 60}, 3163 (1999).

\bibitem{halsey}
T. C. Halsey, Phys. Rev. Lett. {\bf 55}, 1018 (1985); Physica B {\bf 152},
22 (1988).

\bibitem{mychoi}
M. Y. Choi and D. Stroud, Phys. Rev. B {\bf 32}, 7532 (1985); 
Phys. Rev. B {\bf 35}, 7109 (1987);  
J. S. Chung, M. Y. Choi, and D. Stroud, Phys. Rev. B {\bf 38}, 11476 (1988); 
S. Y. Park, M. Y. Choi, B. J. Kim, G. S. Jeon, and J. S. Chung,
Phys. Rev. Lett. {\bf 85}, 3484 (2000). 

\bibitem{doliwa}
B. Doliwa and A. Heuer Phys. Rev. E {\bf 61}, 6898 (2000).


\bibitem{teitel}
S. Teitel, {\em Equilibrium Phase Transitions in
Josephson Junction Arrays} in Proceedings of the Sitges Conference on
Glassy Systems, E. Rubi, Springer, Berlin (1996);
J. P. Straley and G. M. Barnett, Phys. Rev. B {\bf 48}, 3309 (1993).

\bibitem{jrl}
For further details, see
J.-R. Lee and S. Teitel, Phys. Rev. B {\bf 46}, 3247 (1992).

\bibitem{sw}
A. M. Ferrenberg and R. H. Swendsen, Phys. Rev. Lett. {\bf 61}, 2635 (1988);
C. Borgs and R. Koteck\'y, J. Stat. Phys. {\bf 61}, 79 (1990).

\bibitem{config_cg}
S. J. Lee, B. Kim, and J.-R. Lee unpublished.

\bibitem{ediger}
M. Cicerone and M. D. Ediger, J. Chem. Phys. {\bf 104}, 7210 (1996).

\bibitem{sciortino}
F. Sciortino, L. Fabbian, S. H. Chen, and P. Tartaglia, Phys. Rev. E {\bf 56},
5397 (1997).





\end{thebibliography}
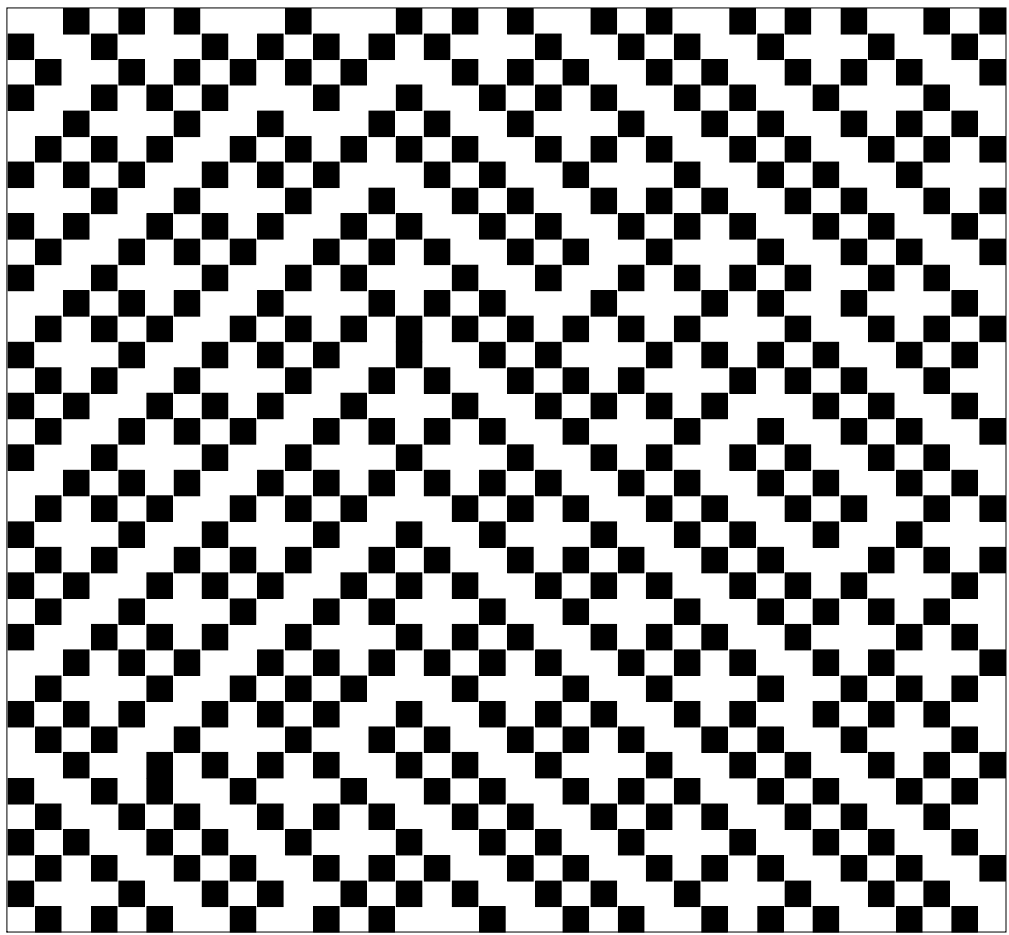
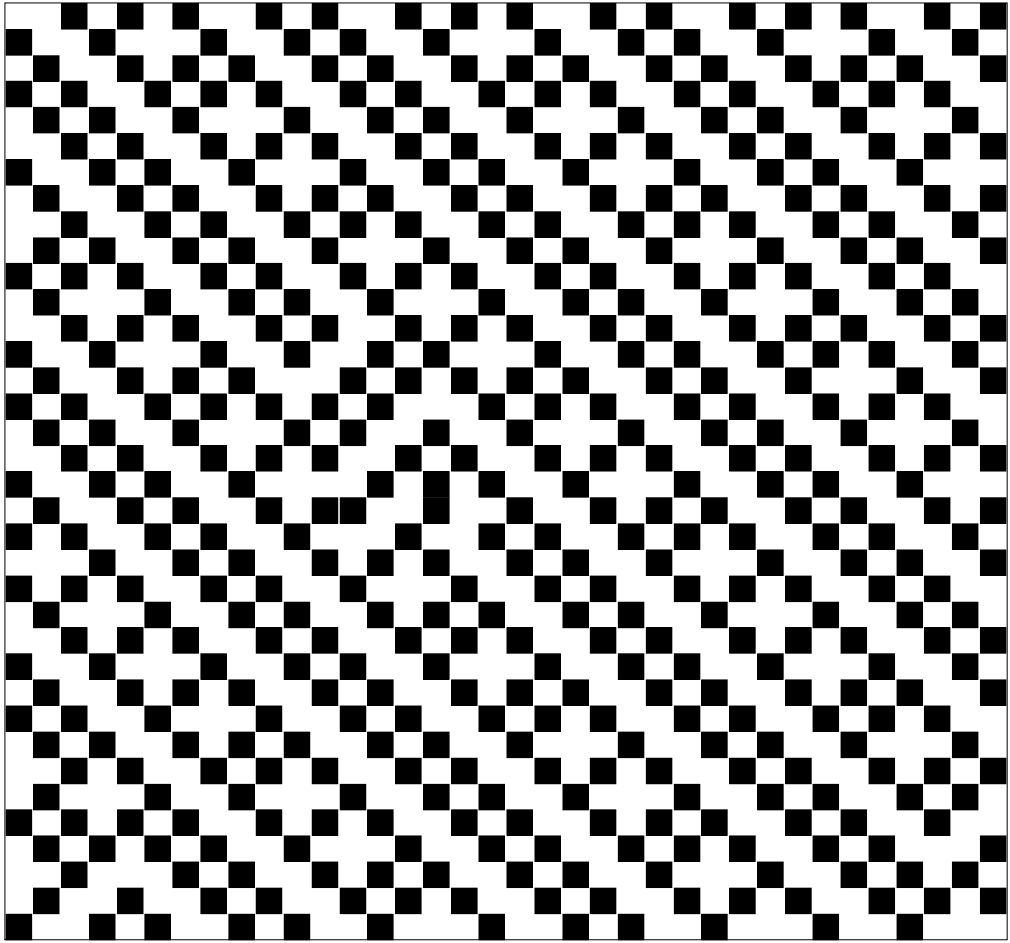
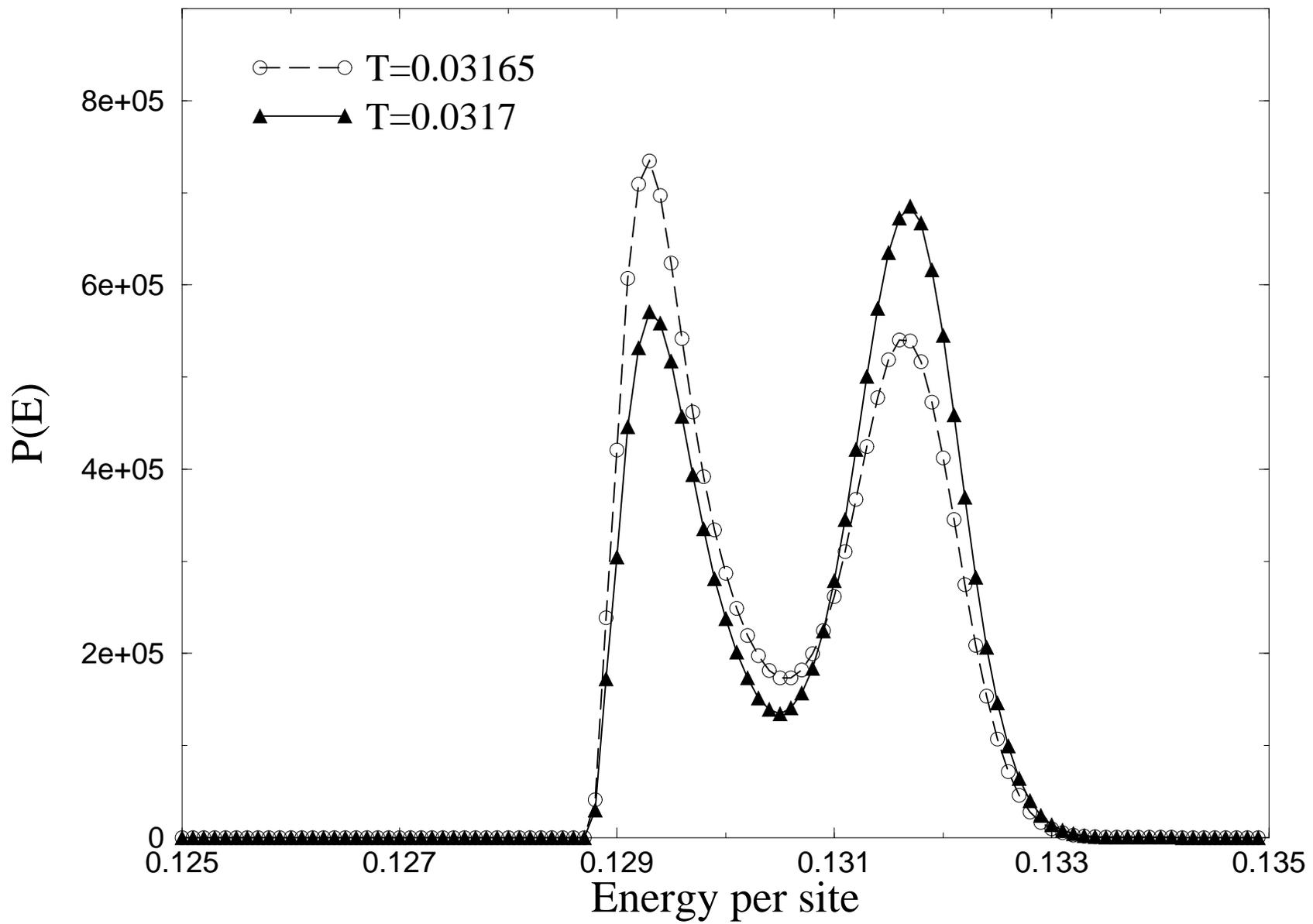
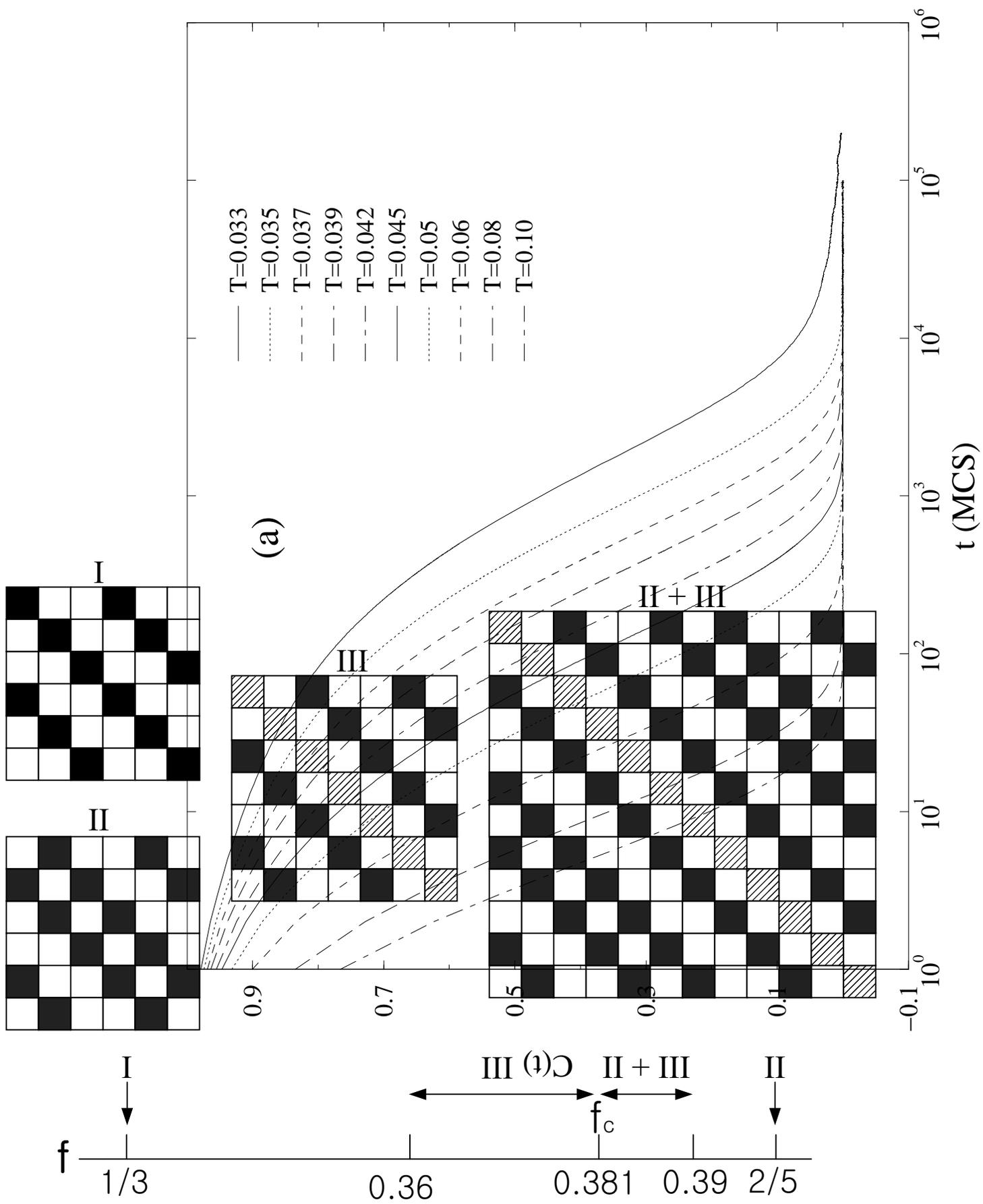
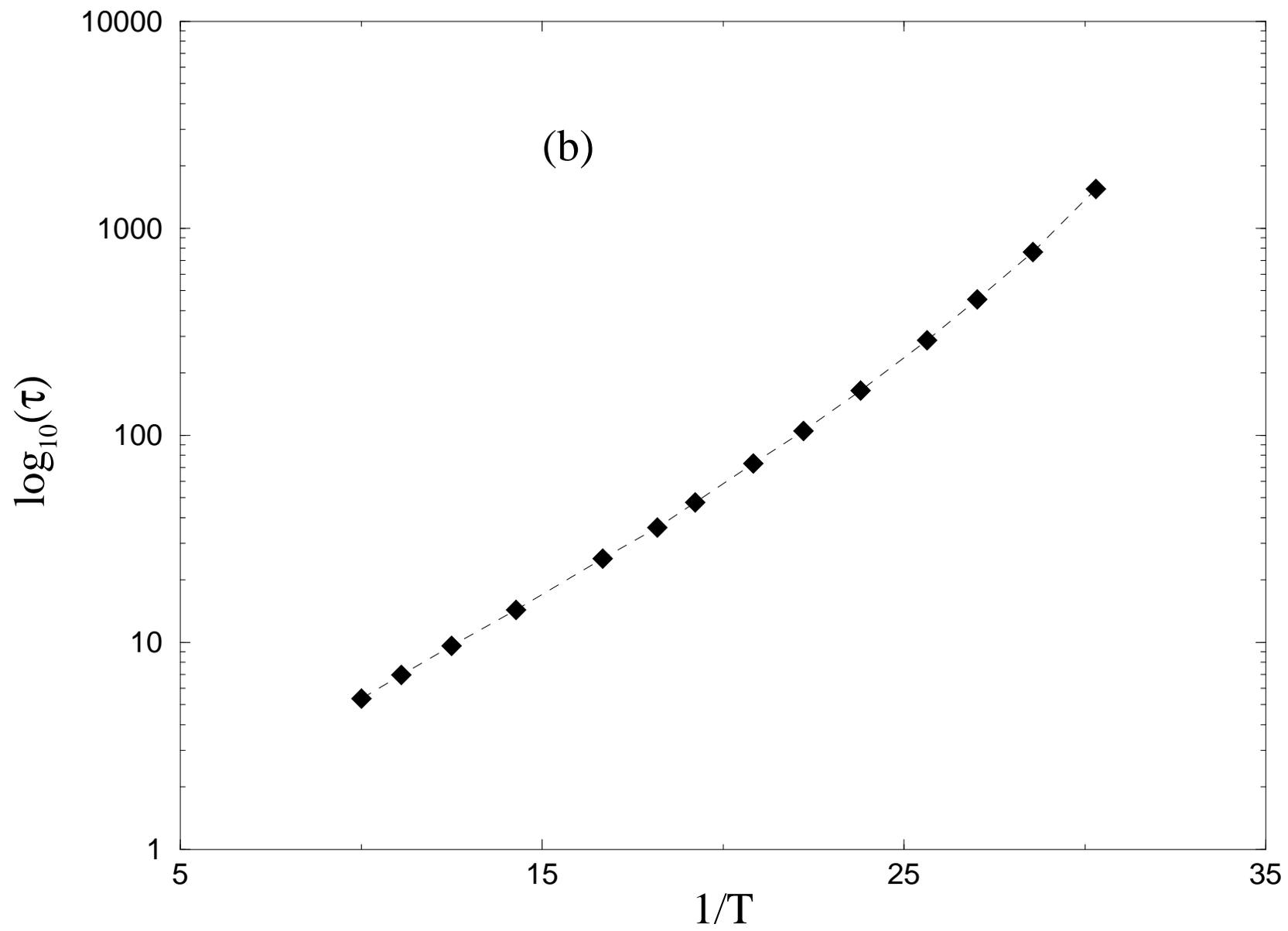
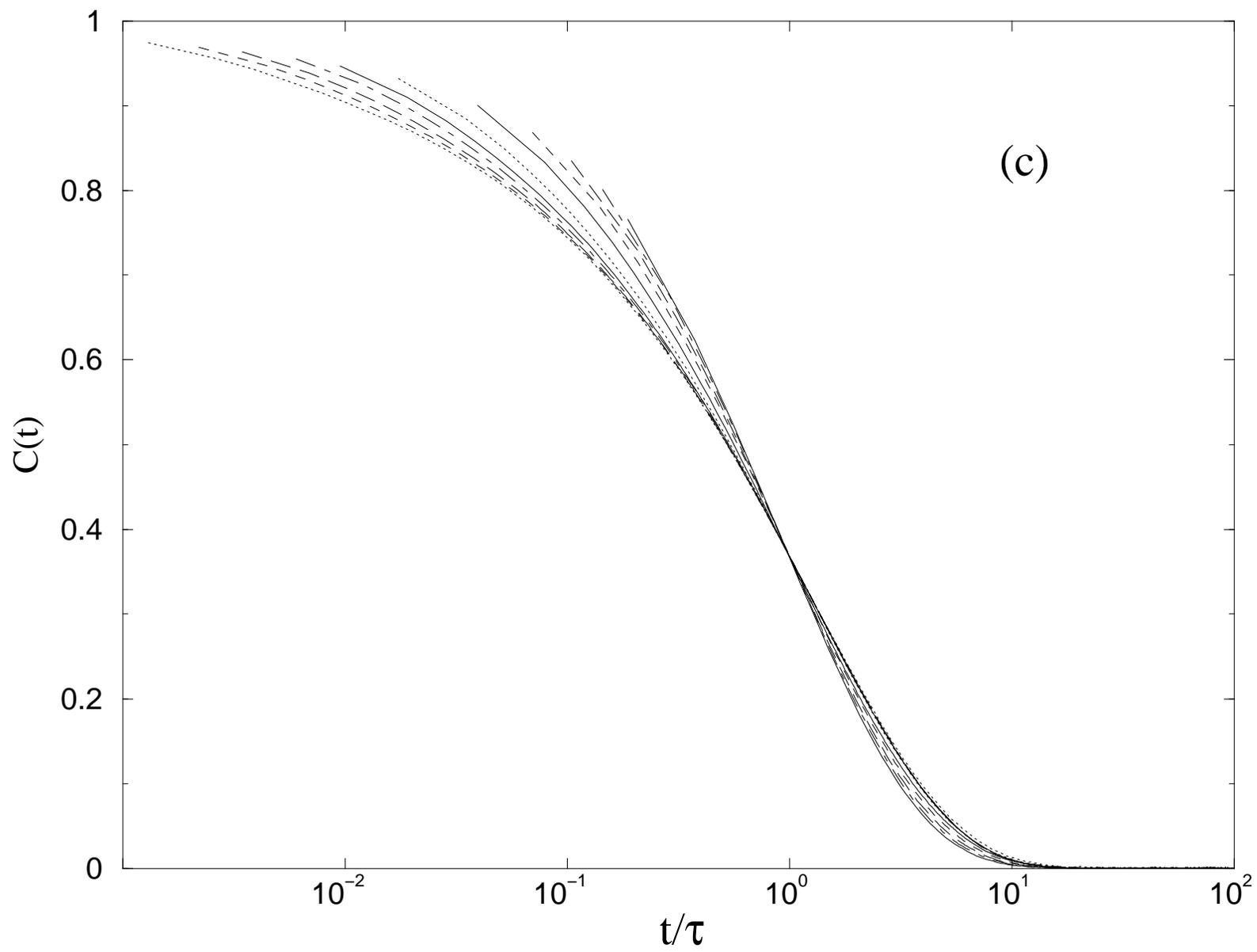
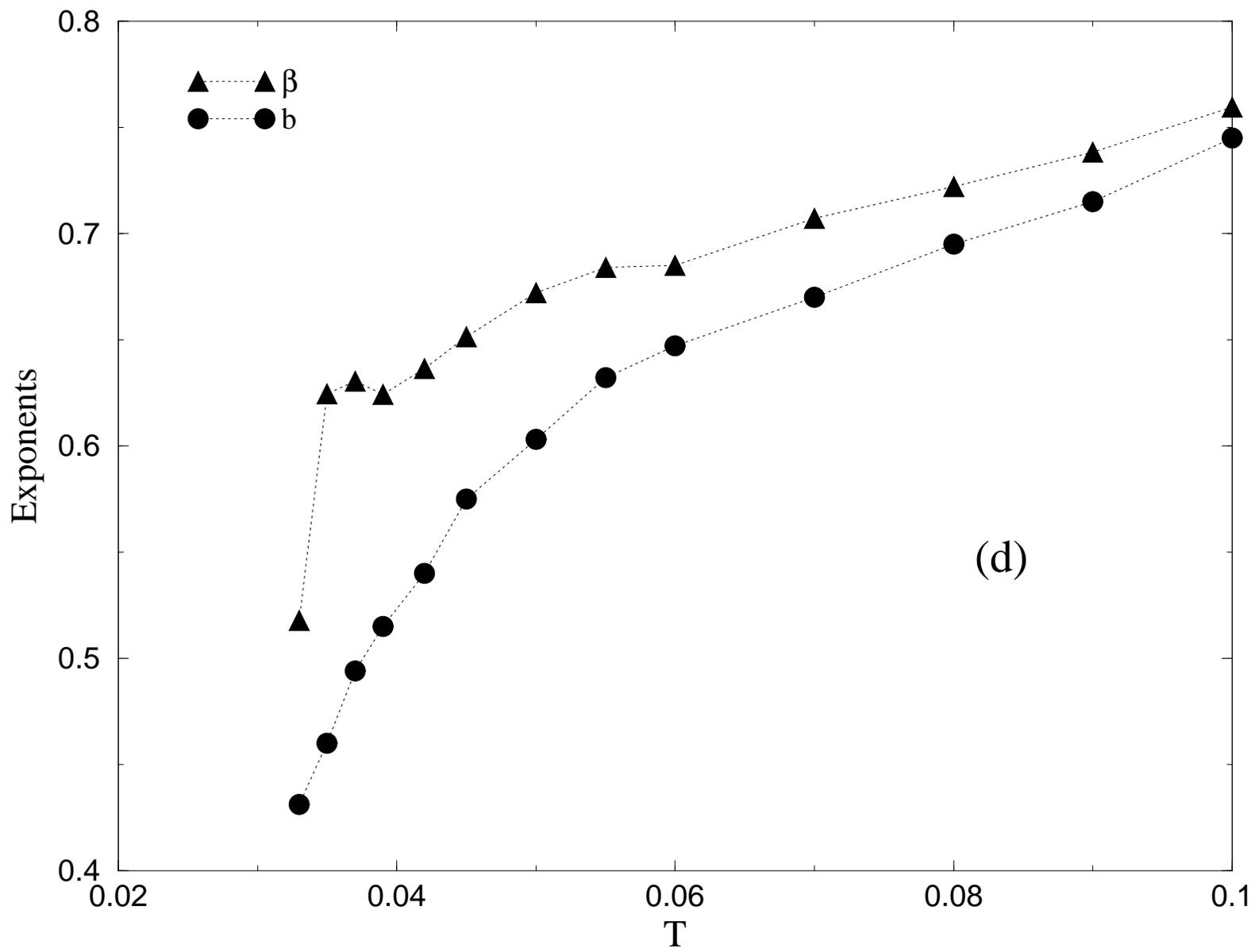
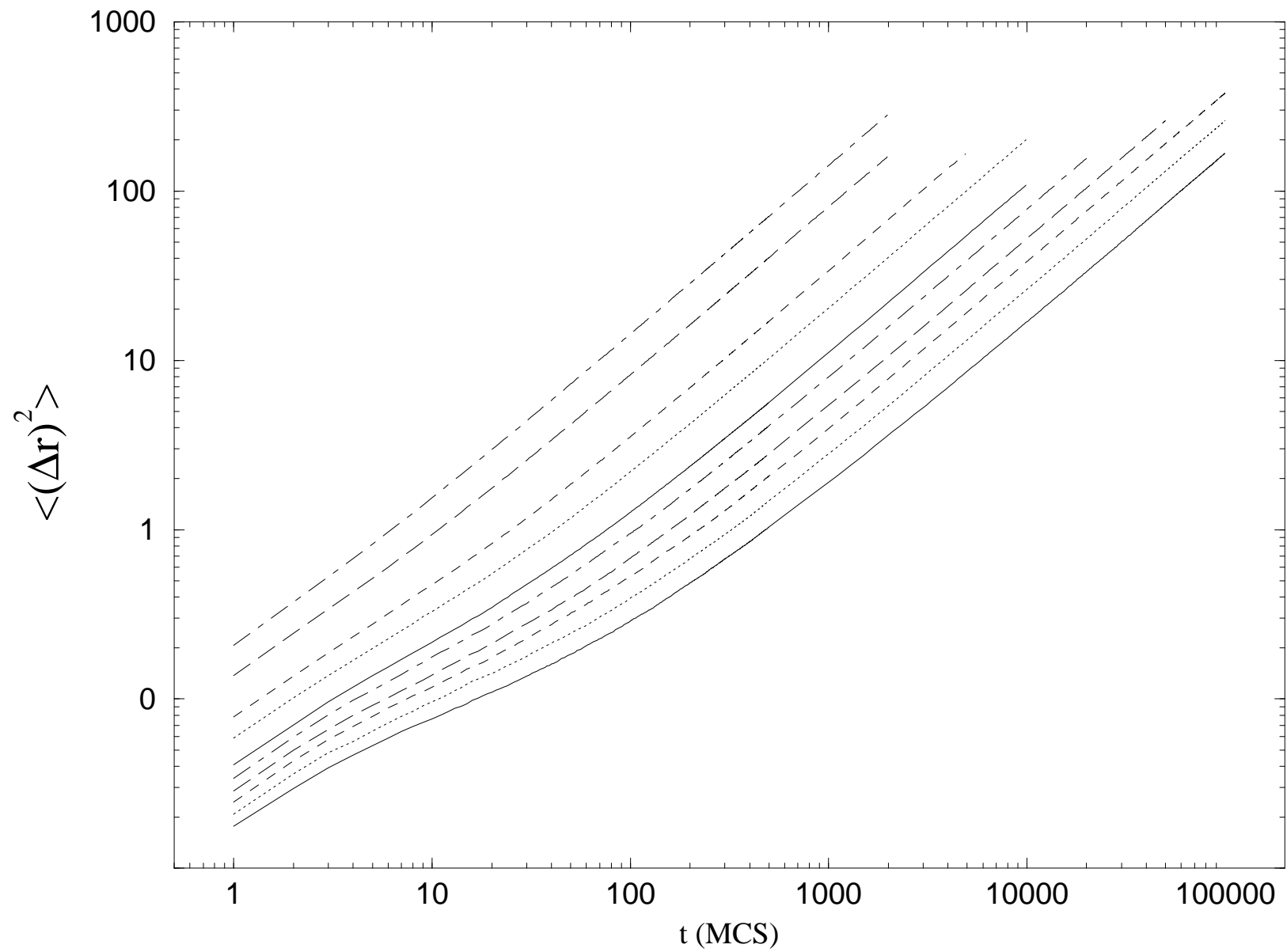
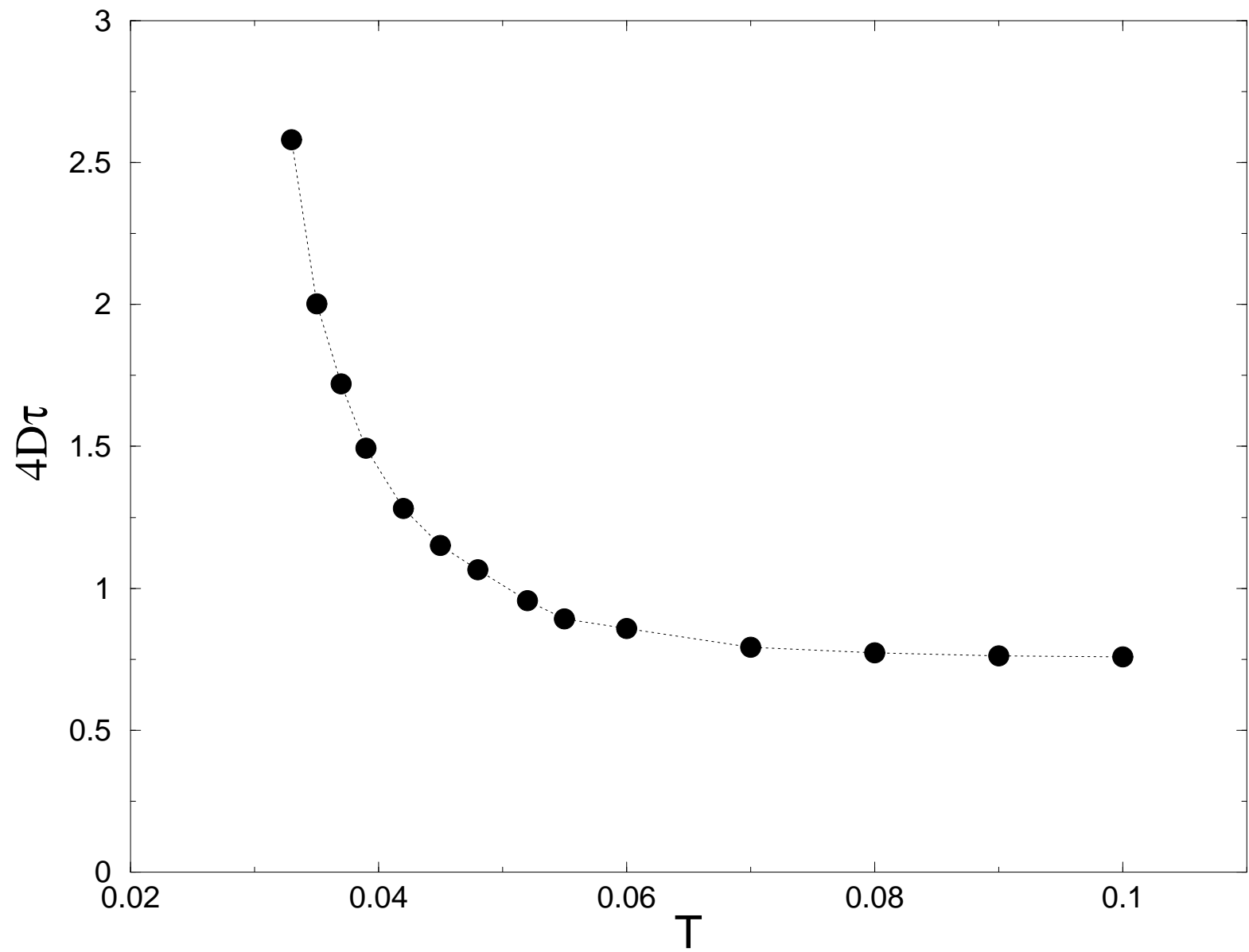
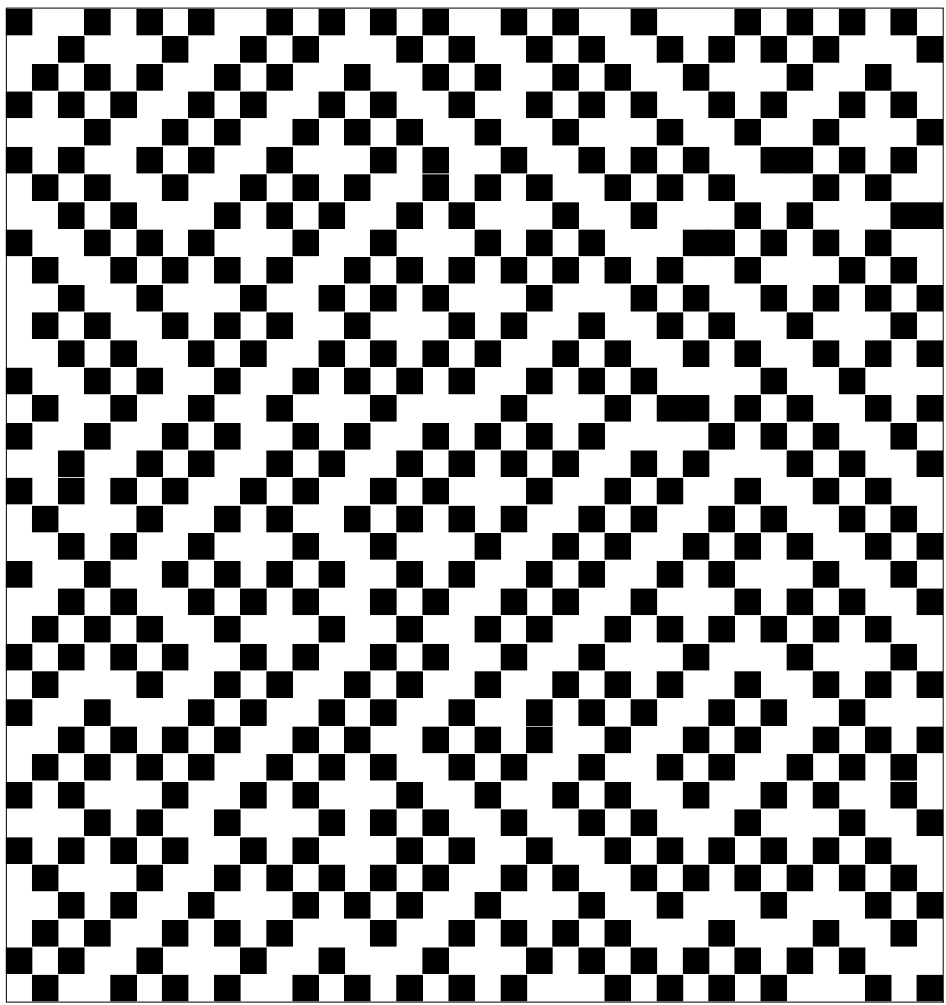
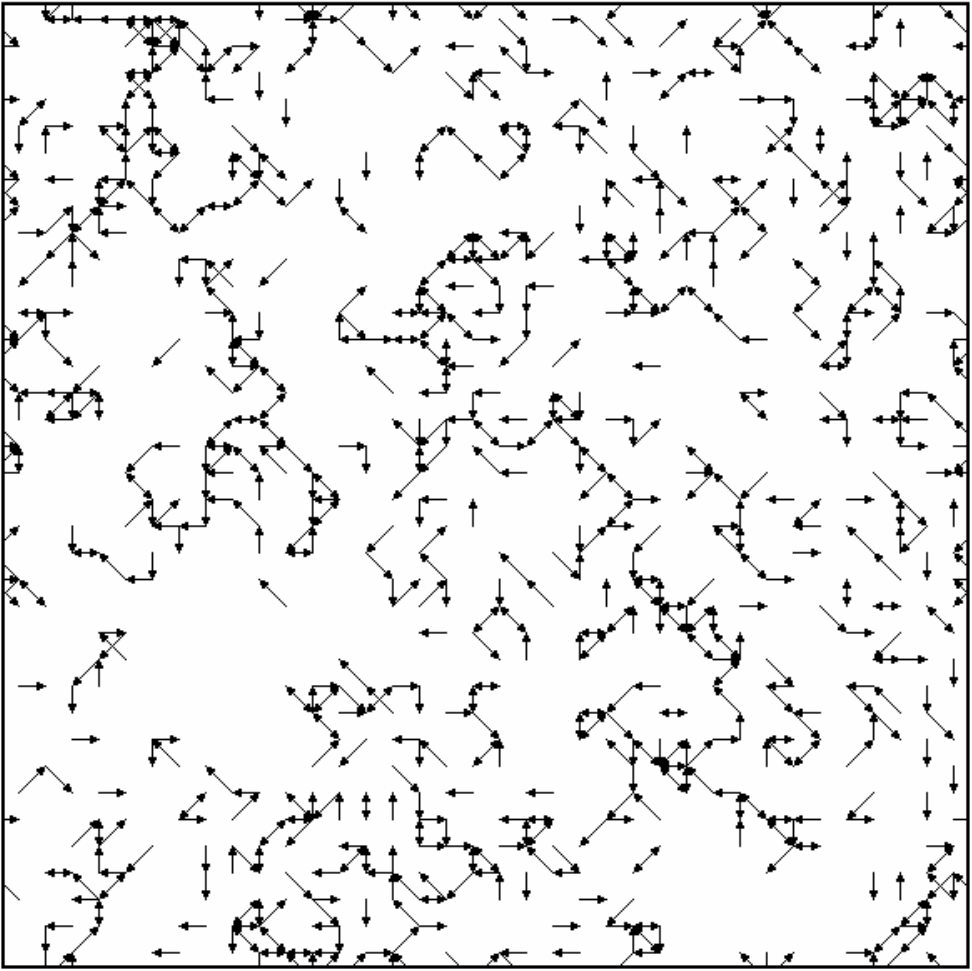
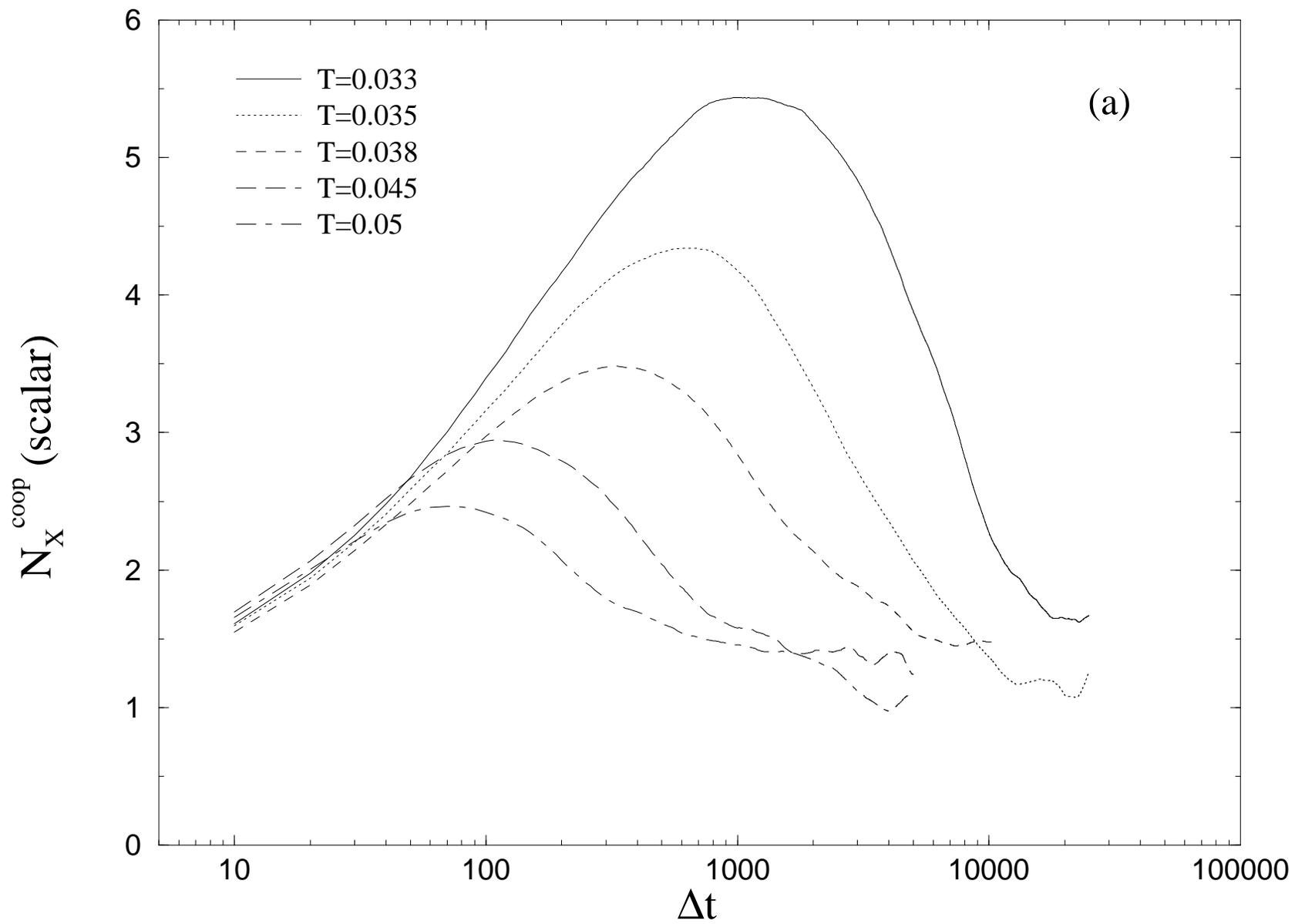
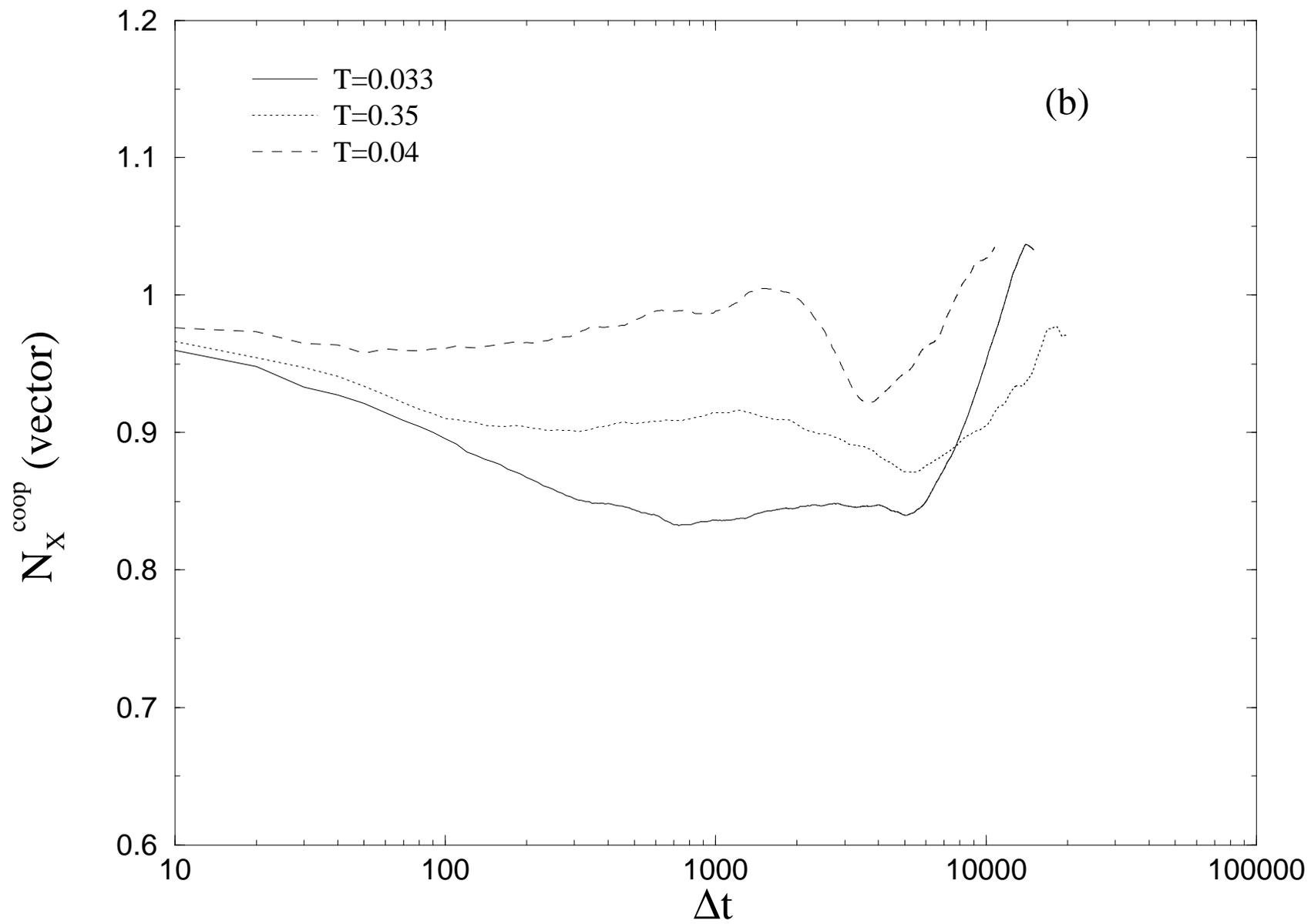
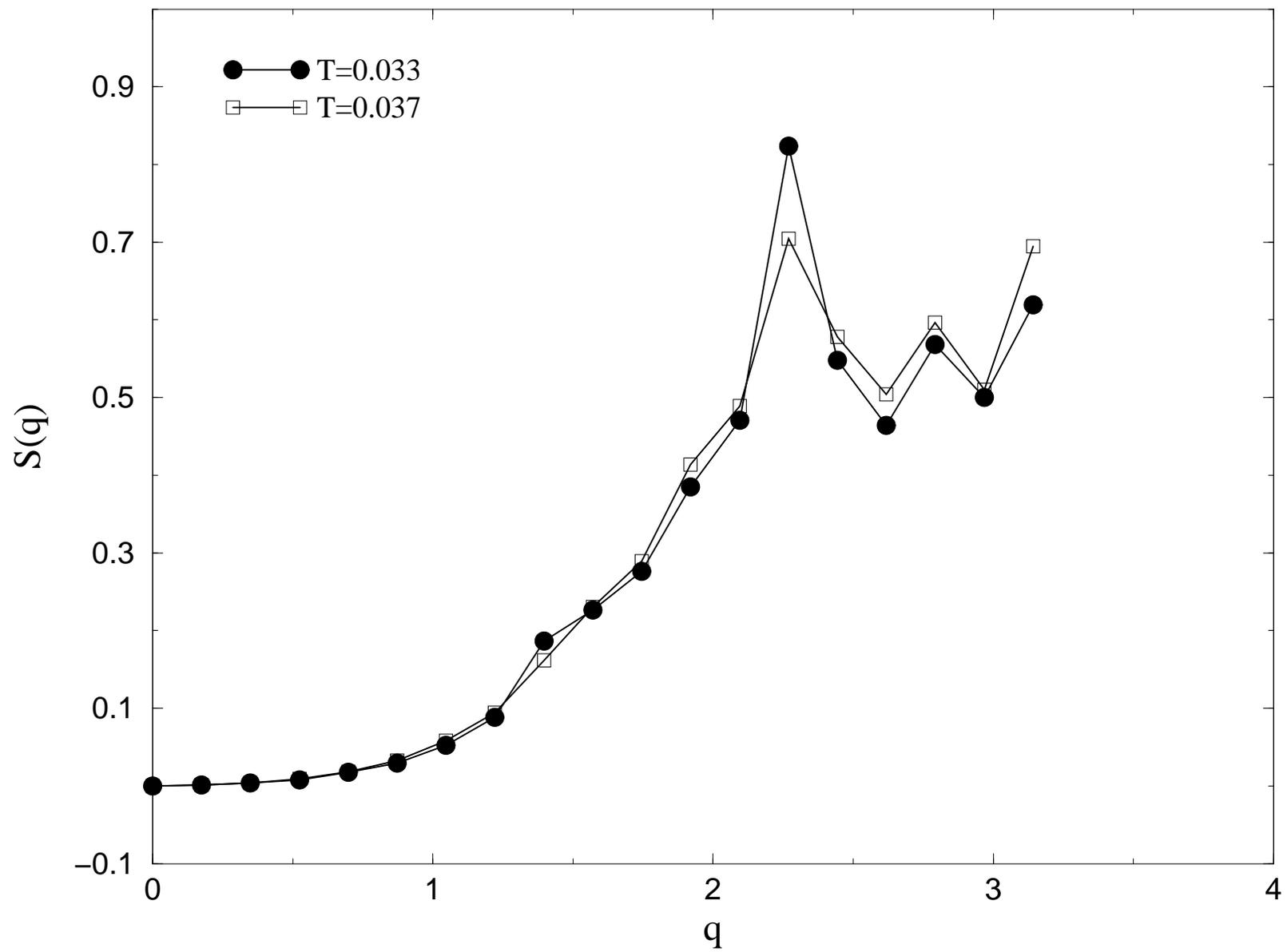
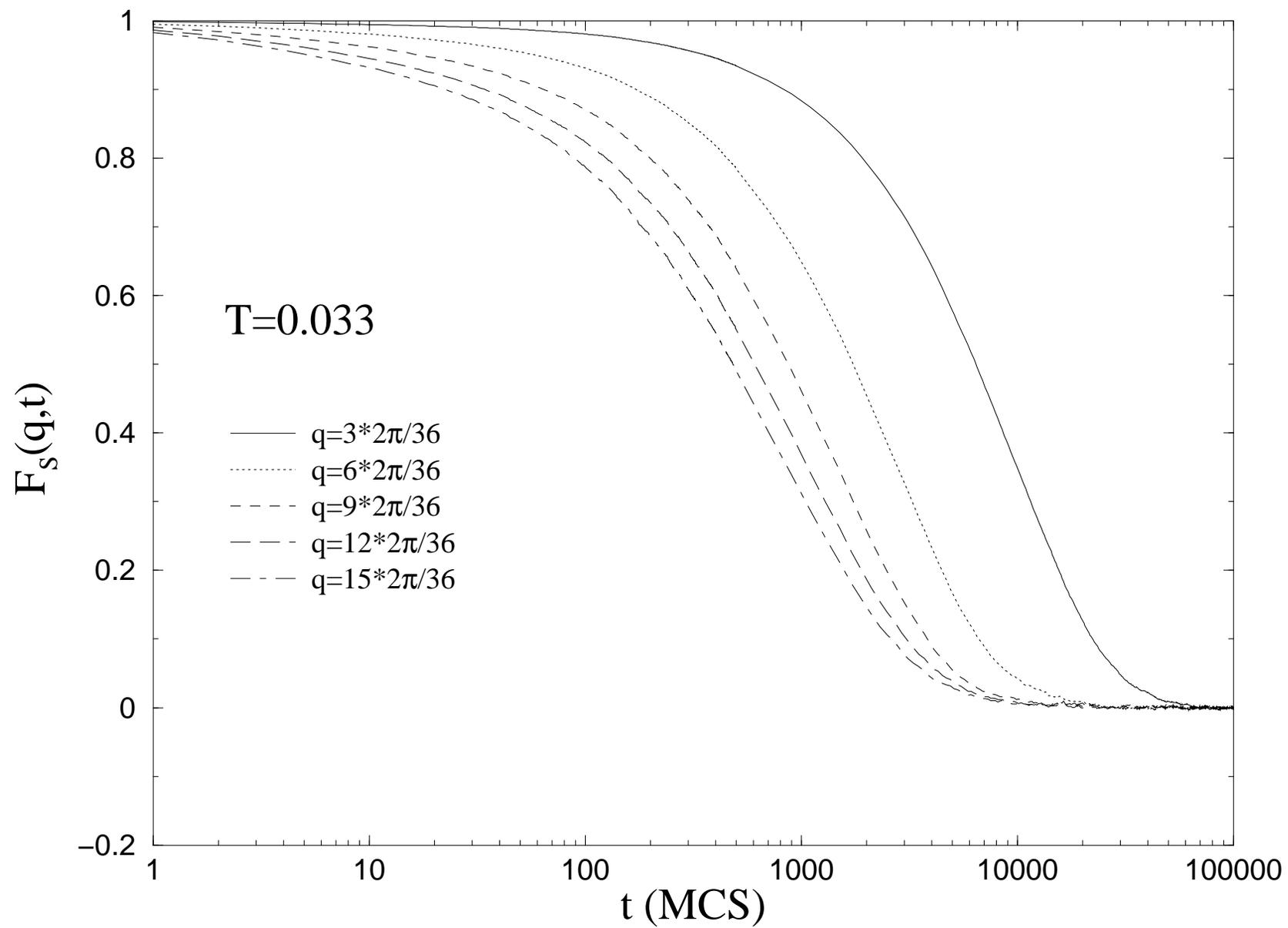
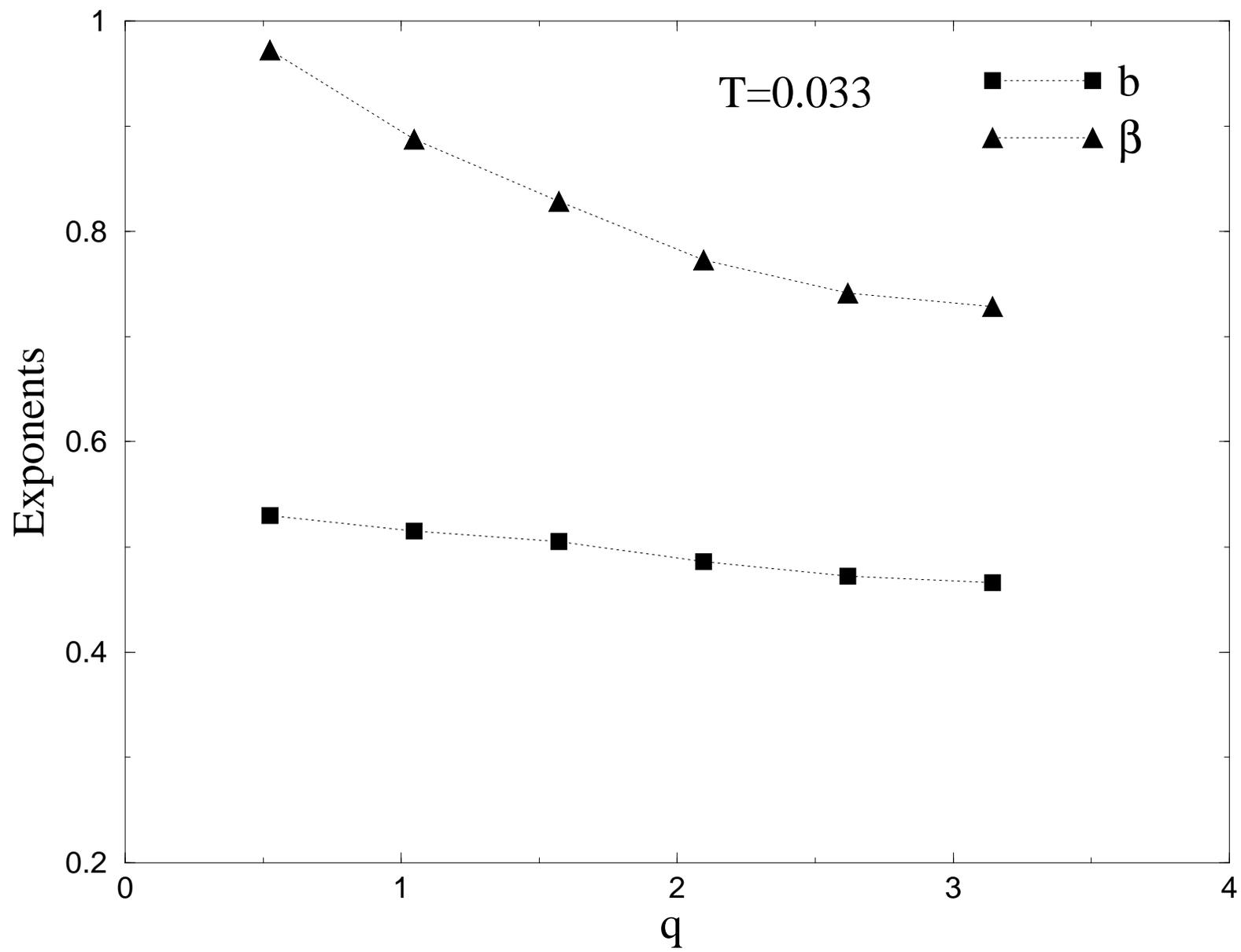
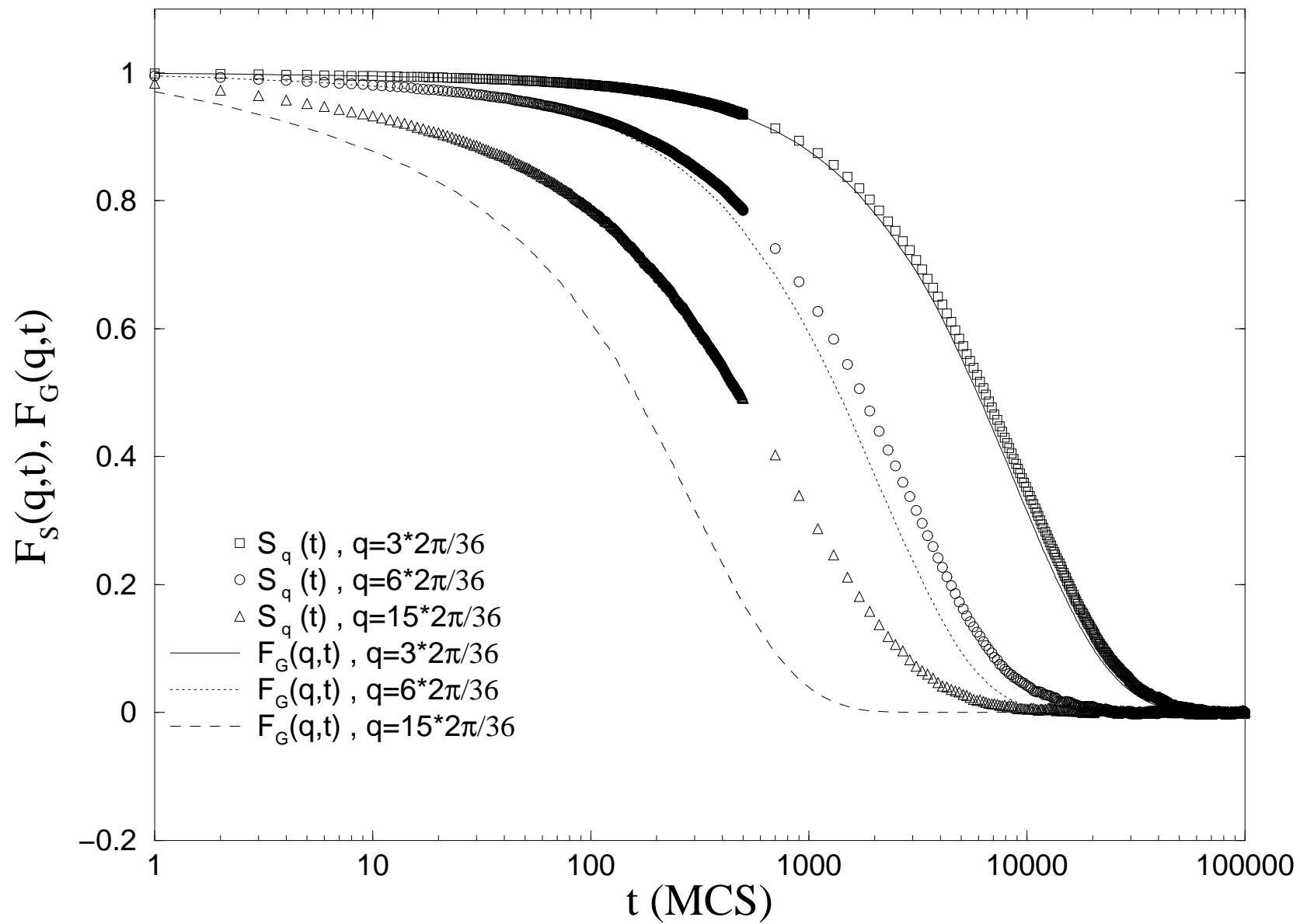
\end{document}